\documentclass{PoS}
\include{graphics}

\title{GRB 080407: an ultra-long burst discovered by the IPN}

\ShortTitle{GRB 080407: an ultra-long burst discovered by the IPN}

\author{\speaker{V. Pal'shin}\\
        Ioffe Physical Technical Institute, St. Petersburg, 194021, Russian Federation\\
        E-mail: \email{val@mail.ioffe.ru}}

\author{K. Hurley\\
       U.C. Berkeley Space Sciences Laboratory, 7 Gauss Way, Berkeley, CA 94720-7450, U.S.A.\\
        E-mail: \email{khurley@ssl.berkeley.edu}}

\author{J. Goldsten\\
       Applied Physics Laboratory, Johns Hopkins University, Laurel, MD 20723, U.S.A.\\
        E-mail: \email{john.goldsten@jhuapl.edu}}

\author{I. G. Mitrofanov\\
       Institute for Space Research, Profsojuznaja 84/32, Moscow 117997, Russian Federation\\
        E-mail: \email{imitrofa@space.ru}}

\author{W. Boynton\\
       Univeristy of Arizona, Lunar and Planetary Laboratory, Tucson, AZ 85721, U.S.A.\\
        E-mail: \email{wboynton@lpl.arizona.edu}}

\author{A. von Kienlin\\
       Max-Planck-Institut f\"{u}r extraterrestrische Physik, Giessenbachstrasse, Garching, 85748 Germany\\
        E-mail: \email{azk@mpe.mpg.de}}

\author{J. Cummings\\
        NASA Goddard Space Flight Center, Code 661, Greenbelt, MD 20771, U.S.A.\\
        E-mail: \email{jayc@milkyway.gsfc.nasa.gov}}

\author{M. Feroci\\
        INAF/IASF-Roma, via Fosso del Cavaliere 100, 00133, Roma, Italy\\
        E-mail: \email{Marco.Feroci@iasf-roma.inaf.it}}

\author{R. Aptekar, D. Frederiks, S. Golenetskii, E. Mazets, D. Svinkin\\
        Ioffe Physical Technical Institute, St. Petersburg, 194021, Russian Federation\\
        E-mail: \email{aptekar@mail.ioffe.ru}, \email{fred@mail.ioffe.ru}, \email{golen@mail.ioffe.ru}, \email{mazets@mail.ioffe.ru}, \email{svinkin@mail.ioffe.ru}}

\author{D. Golovin, M. L. Litvak, A. B. Sanin\\
       Institute for Space Research, Profsojuznaja 84/32, Moscow 117997, Russian Federation\\
        E-mail: \email{dimamsu@mail.ru}, \email{max@cgrsmx.iki.rssi.ru}, \email{sanin@mx.iki.rssi.ru}}

\author{C. Fellows, K. Harshman\\
       Univeristy of Arizona, Lunar and Planetary Laboratory, Tucson, AZ 85721, U.S.A.\\
        E-mail: \email{cfellows@lpl.arizona.edu}, \email{karl@lpl.arizona.edu}}

\author{R. Starr\\
       Physics Department, The Catholic University of America, Washington, DC 20064, USA\\
        E-mail: \email{richard.starr@gsfc.nasa.gov}}

\author{A. Rau, X. Zhang\\
       Max-Planck-Institut f\"{u}r extraterrestrische Physik, Giessenbachstrasse, Garching, 85748 Germany\\
        E-mail: \email{arau@mpe.mpg.de}, \email{zhangx@mpe.mpg.de}}

\author{V. Savchenko\\
        ISDC Data Centre for Astrophysics, Ch. d'Ecogia 16, 1290, Versoix, Switzerland\\
        E-mail: \email{Volodymyr.Savchenko@unige.ch}}

\author{S. D. Barthelmy, N. Gehrels, H. Krimm, D. Palmer\\
        NASA Goddard Space Flight Center, Code 661, Greenbelt, MD 20771, U.S.A.\\
        E-mail: \email{scott@milkyway.gsfc.nasa.gov}, \email{gehrels@lheavx.gsfc.nasa.gov}, \email{krimm@milkyway.gsfc.nasa.gov}, \email{palmer@lanl.gov}}

\author{E. Del Monte, M. Marisaldi\\
        INAF/IASF-Roma, via Fosso del Cavaliere 100, 00133, Roma, Italy\\
        E-mail: \email{delmonte@iasf-roma.inaf.it}, \email{marisaldi@iasfbo.inaf.it}}

\abstract{We present observations of the extremely long GRB 080704 obtained with the instruments of the Interplanetary Network (IPN). The observations reveal two distinct emission episodes, separated by a $\sim$1500~s long period of quiescence. The total burst duration is about 2100~s. We compare the temporal and spectral characteristics of this burst with those obtained for other ultra-long GRBs and discuss these characteristics in the context of different models.}

\FullConference{Gamma-Ray Bursts 2012 Conference -GRB2012,\\
        May 07-11, 2012\\
        Munich, Germany}

\begin{document}

\section{Observations}
\begin{figure}[h!]
\includegraphics[width=0.4\textwidth]{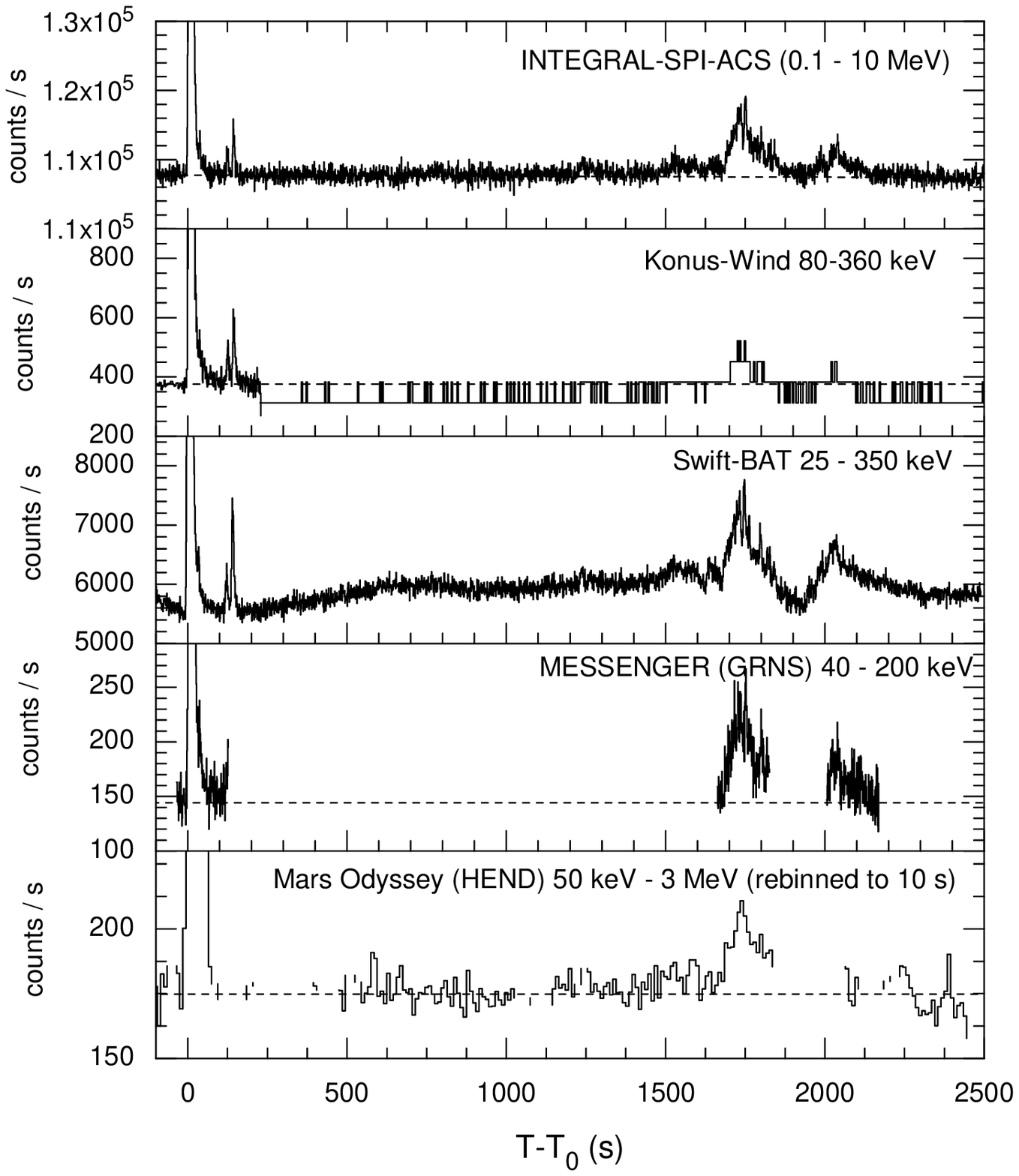}
\hfill
\includegraphics[width=0.4\textwidth]{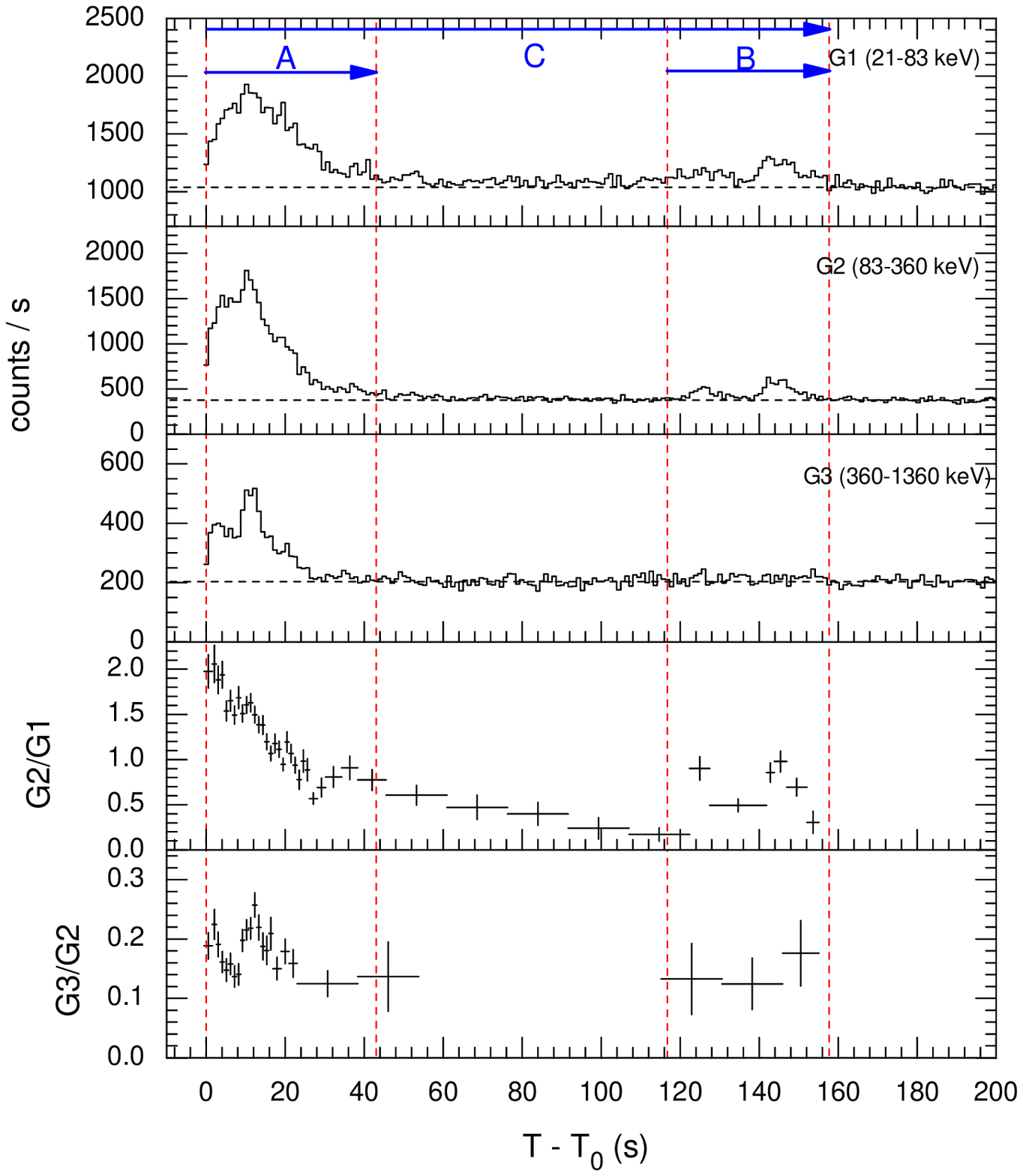}
\caption{GRB 080407 light curve. \emph{Left:} full light curve as observed by different IPN instruments. T$_0$= T$_0^*$(KW) = 74529.471 s UT (20:42:09.471) (* -- corrected for the propagation time delay to Earth; T$_0$(KW) -- the Konus-Wind trigger time). Numerous noise spikes  have been removed from the HEND lc. \emph{Right:} first emission episode in three energy bands and hardness ratios as observed by Konus-Wind.}
\label{figGRBlc}
\vspace{-0.6cm}
\end{figure}

The first emission episode of GRB 080704  was detected at about 74529~s UT (20:42:09) on July 4th 2008.
It displays a bright initial pulse followed by two weaker pulses -- see Fig.~\ref{figGRBlc}. The total duration of the episode is $\sim$160~s (T$_{90} =142.2 \pm 1.5$ s in the 21--1360 keV band\footnote{as observed by  Konus-Wind}). The whole episode was observed by Konus-Wind, INTEGRAL-SPI-ACS, and Swift-BAT (outside of the coded FoV). The initial pulse was also detected by AGILE (SuperAGILE \& MCAL), Mars Odyssey (GRS \& HEND), and MESSENGER (GRNS).

The second episode was detected at about 76000~s UT (21:06:40), i.e. $\sim$1500~s after the onset of the burst. It shows two broad pulses with a total duration of $\sim$400~s and a weaker emission before the pulses -- see Fig.~\ref{figGRBlc}. Both pulses were  detected by INTEGRAL-SPI-ACS, Swift-BAT (outside of the coded FoV), and MESSENGER (GRNS). 
The first pulse is also detected in the HEND data, however, the time interval of the second pulse is strongly affected by noise spikes. Unfortunately this episode was seen by Konus-Wind only in the housekeeping data (80--360 keV range, 3.68~s resolution) due to data readout after the trigger on the initial pulse, so no spectral data are available for this episode.

\section{IPN localization}
\begin{figure}[t!]
\includegraphics[width=0.4\textwidth]{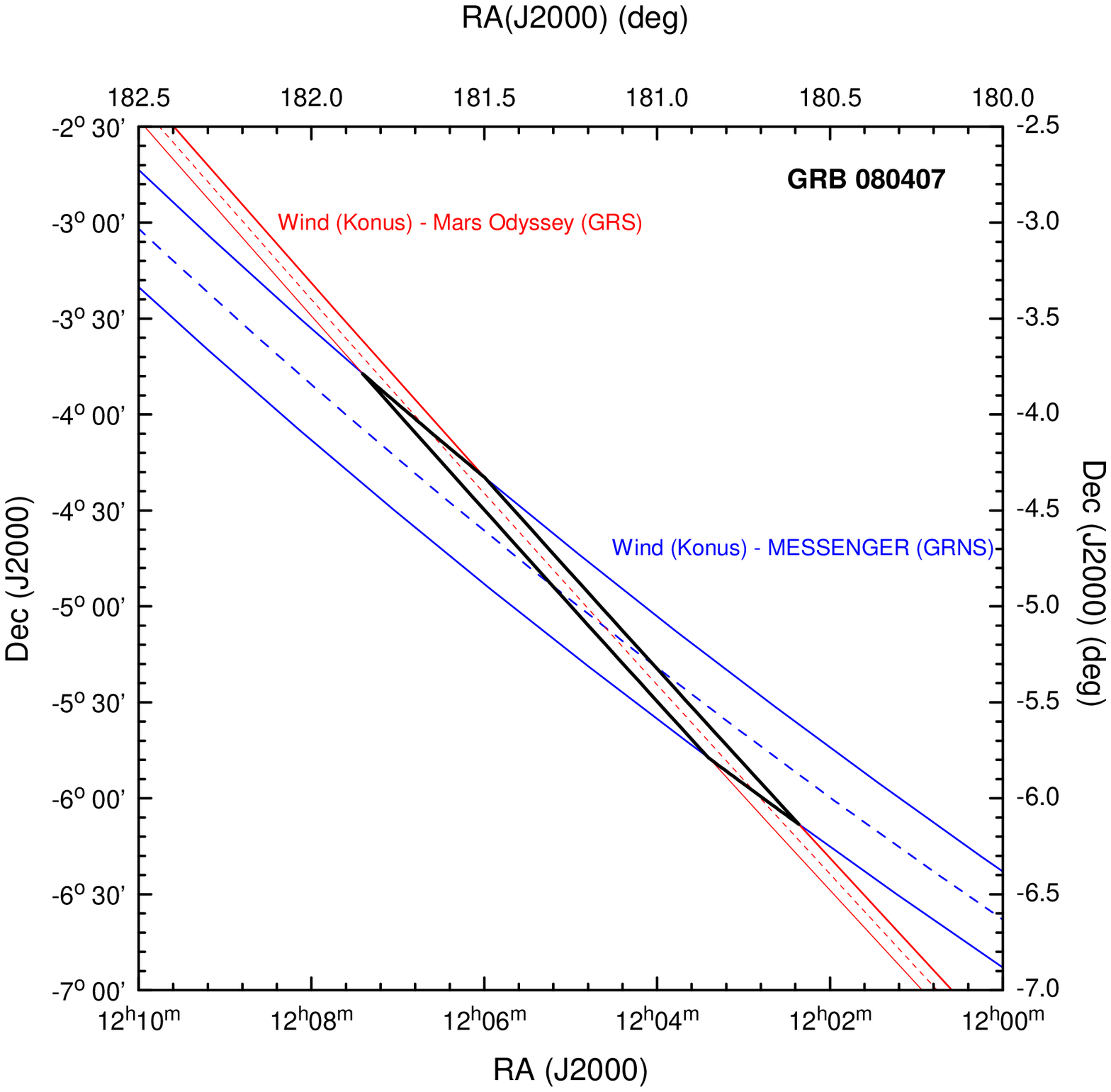}
\hfill
\includegraphics[width=0.4\textwidth]{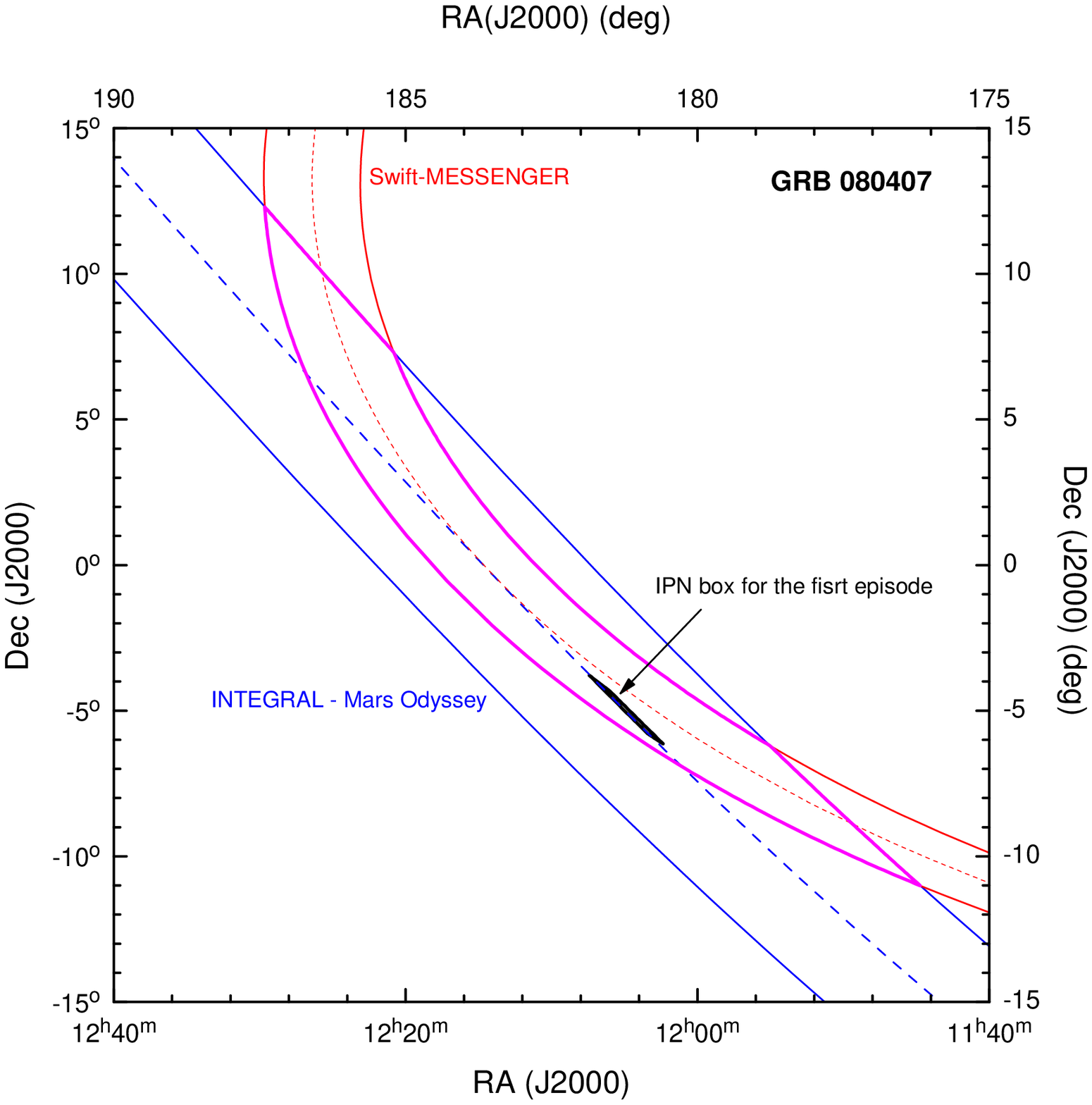}
%
%
%
%
\label{figIPNbox}
\caption{IPN error boxes for GRB 080407. \emph{Left:} for the first episode. \emph{Right:} for the second episode.}
\vspace{-0.5cm}
\end{figure}

We have triangulated the first episode to the rather small 3 sigma error box shown in Fig.~\ref{figIPNbox} (left panel). The error box area is 582 sq. arcmin, and its maximum dimension is 2.67 deg. For the second episode we have derived the rather large 3 sigma error box shown in Fig.~\ref{figIPNbox} (right panel). The error box area is 17.1 sq. deg. This box contains the small box for the first episode. The coordinates of the boxes are given in Table~\ref{TableIPNbox}.

To estimate the probability of chance coincidence we can multiply the relative area of the large error box (that is $\Omega/4 \pi$ sr = 17.1 sq. deg/ 41252 sq. deg = $4.1 \times 10^{-4}$) by the probability to have two bursts in $t \lesssim 1500$ s (that is, for the 2008 IPN detection rate, R, of $\sim$250 GRBs/year, $p = 1 - \exp(-Rt) \simeq 0.012$) to get $\sim 5 \times 10^{-6}$. Multiplying this number by the total number of  GRBs detected by the 3rd IPN so far ($\sim$ 6000), we get a 3\% probability of observing such a close (in time and on the sky) pair of IPN events for the entire 3rd IPN history. Actually this probability is significantly overestimated since we counted all IPN bursts, rather than just those bright enough to be detected by the least sensitive instrument in the IPN, MO HEND.

This simple estimate shows that these episodes likely belong to the same ultra-long GRB.

\begin{table}
{
\caption{IPN localization of GRB 080407}
\fontsize{8pt}{10pt}\selectfont
\begin{center}
\begin{tabular}{lccccr}
\hline
&\multicolumn{2}{c}{first episode}&\hspace{3mm}&\multicolumn{2}{c}{second episode}\\
\hline
& RA(2000), deg & Dec(2000), deg &&  RA(2000), deg & Dec(2000), deg\\
\hline
Center & 181.151 (12h 04m 36s) & -5.110 ( -5d 06' 34")&& & \\
Corner1 & 180.589 (12h 02m 21s)  & -6.136 (-6d 08' 11") &&  187.413 (12h 29m 39s) & +12.295 (+12d 17' 42")\\
Corner2 & 180.851 (12h 03m 24s)  & -5.789 (-5d 47' 22") &&  185.204 (12h 20m 49s) & +7.311 ( +7d 18' 39")\\
Corner3 & 181.852 (12h 07m 24s)  & -3.787 (-3d 47' 13") &&  178.760 (11h 55m 02s) & -6.199 ( -6d 11' 56")\\
Corner4 & 181.499 (12h 06m 00s)  & -4.327 (-4d 19' 38") && 176.185 (11h 44m 44s)  & -11.005 (-11d 00' 17")\\
\hline
\end{tabular}
\end{center}
\label{TableIPNbox}
\vspace{-0.8cm}
}
\end{table}

\vspace{-0.5cm}

\section{Spectral analysis and energetics}
The Konus-Wind spectral data from 20 to 1200 keV for the first episode were fitted by the GRBM (Band) model for three specific time intervals denoted on Figure~\ref{figGRBlc} (right panel). The results are given in Table~\ref{TableGRBsppar}. All spectra have a high-energy power-law tail with $\beta \sim -2.3$. The spectrum of the weaker pulses at $\sim$T$_0$+117~s is significantly softer (in terms of $E_{peak}$) than the spectrum of the initial bright pulse.

Using the parameters of the time-integrated spectrum (Int. C), the calculated fluence of this episode is $(1.43 \pm 0.04)\times 10^{-4}$~erg~cm$^{-2}$, and the peak flux is $(7.76 \pm 0.38)\times 10^{-6}$~erg~cm$^{-2}$~s$^{-1}$ as measured from 10.4~s on a 1~s timescale (both in the 20--1000 keV)\footnote{in the 20 keV - 10 MeV band they are $(2.01 \pm 0.27)\times10^{-4}$~erg~cm$^{-2}$ and $(1.08 \pm 0.10)\times 10^{-5}$~erg~cm$^{-2}$~s$^{-1}$.}.

There are no spectral data for the second episode. Comparison of the counts accumulated over several time intervals by different instruments shows that the hardness of the second episode is closer to the hardness of the weaker pulses of the first episode. Using a conversion factor of 1 SPI-ACS count = $(4.1 \pm 0.5)\times10^{-10}$ erg cm$^{-2}$ (obtained for the weaker pulses of the first episode; for the initial pulse this factor is $(2.31 \pm 0.04)\times 10^{-10}$~erg~cm$^{-2}$~cnt$^{-1}$), the estimated fluence of this episode is $\sim 3 \times 10^{-4}$~erg~cm$^{-2}$. So the total fluence of GRB 080704 is about $4.4 \times 10^{-4}$~erg~cm$^{-2}$.

\begin{table}
{
\caption{Spectral parameters of GRB 080407 for three time
intervals denoted on Fig.~1 (right panel). All errors are at 90\% CL.}
\fontsize{8pt}{10pt}\selectfont
\begin{center}
\begin{tabular}{crrlclc}
\hline
Int & Tstart (s) & dT (s) & $\alpha$ &  $E_{peak}$ (keV) & $\beta$ & $\chi^2$/dof (null hypothesis probability) \\
\hline
A   & 0 & 43 &  -1.02$\pm$0.06 &    325$_{-25}^{+29}$ & -2.43$_{-0.27}^{+0.16}$ &   60.8/59 (0.41)\\
B   & 117   & 41    & -1.49$_{-0.37}^{+0.76}$   & 114$_{-44}^{+77}$ & -2.25(<-2.02) &   48.3/59 (0.84)\\
C (total)   & 0 & 158   & -1.15$_{-0.09}^{+0.10}$   & 287$_{-35}^{+42}$ & -2.35$_{-0.40}^{+0.20}$ & 57.1/59 (0.54)\\
\hline
\end{tabular}
\end{center}
\label{TableGRBsppar}
\vspace{-0.8cm}
}
\end{table}

\vspace{-0.5cm}

\section{Comparison with other ultra-long GRBs}
Table~\ref{TableULGRBs} contains a comparison of the parameters of the six classical gamma-ray bursts with durations >1000 s. It does not include very long underluminous XRFs like XRF 060218, which are thought to be a different phenomenon from classical GRBs. The table also does not list several ultra-long GRB candidates found in the BATSE data \cite{Tikhomirova2005}, since it is difficult to establish reliably that the various emission episodes belong to the same GRB.

Only a short note on GRB 840304 has been published, so it is not quite clear whether it really belongs to the class of ultra-long GRBs and has a very large fluence.

In common these ultra-long GRBs have large fluences (except GRB 020410 which has rather moderate fluence) due to their very long durations and rather hard spectra with $E_{peak}$ of hundreds of keV. But they have quite different morphologies: GRB 971208 and GRB 060814B shows a single, smooth, FRED-like pulse, whereas GRB 020410, GRB 080407, and GRB 091024 show several emission episodes separated by long period(s) of quiescence.

There is also a difference in the light curves of GRB 080407 and two other multi-episode bursts: GRB 080407 shows a bright initial pulse followed by substantially weaker pulses, whereas GRB 020410 and GRB 091024 do not display a decrease of pulse intensities with time (on the contrary, the last pulse of GRB 091024 is brighter that the first one). But in all three cases the initial pulse is significantly harder than subsequent ones.

\begin{table}
{
\caption{Ultra-long GRBs}
\fontsize{8pt}{10pt}\selectfont
\begin{center}
\begin{tabular}{lp{1.5cm}ccp{3cm}cp{1.5cm}c}
\hline
GRB & Tstart &  dT  & Fluence & lc shape &  z   & $E_{iso}$ &   Refs.\\
    & s UT &    s   & erg cm$^{-2}$ & & & erg & \\
\hline
840304  & ? & $\sim$1200 &  $\sim 2.8 \times 10^{-3}$ & Two broad pulses ($\sim$200 s) + extended tail ($\sim$1000 s) & -    & $\sim 7.6 \times 10^{54}$ (for z=1)   & \cite{Klebesadel1984}\\
971208  & 28092 (07:48:12)  & $\sim$2500 &  $(2.55 \pm 0.11) \times 10^{-4}$ &  Single smooth FRED-like pulse   &  - & $\sim 6.9 \times 10^{53}$ (for z=1) &    \cite{IAUC6785,Giblin2002,Palshin2008}\\
020410  & 38380 (10:39:40) &    $\sim$1600  & $\sim 2.8 \times 10^{-5}$ &   Multi-episode & $\simeq$0.5 &   $\sim 1.8 \times 10^{52}$ & \cite{Nicastro2004,Levan2005}\\
060814B & 37070 (10:17:50)  & $\sim2700$ &  $(2.35\pm0.22)\times10^{-4}$ &  Single smooth FRED-like pulse & - & $\sim 6.4\times 10^{53}$ (for z=1) &    \cite{Palshin2008}\\
080407 &    74529 (20:42:09)    & $\sim$2100 &  $\sim 4.5\times10^{-4}$ &   Multi-episode   & - & $\sim 1.2 \times10^{54}$ (for z=1) & this work\\
091024  & 32161 (08:56:01)  & $\sim$1200 & $(1.13 \pm 0.12) \times 10^{-4}$ & Multi-episode & 1.09  & $\sim 3.5 \times 10^{53}$ &    \cite{Marshall2009,Golenetskii2009,Gruber2011}\\
\hline
\end{tabular}
\end{center}
\label{TableULGRBs}
\vspace{-0.8cm}
}
\end{table}

\section{Summary}
Only a few ultra-long GRBs (with durations > 1000 s) have been reported to date.
We have presented observations of GRB 080407, probably the longest multi-episode GRB detected so far. The measured burst fluence of $\sim 4 \times 10^{-4}$ erg cm$^{-2}$ (20--1000 keV) is among the largest observed in long GRBs.
This burst demonstrates similarities with other ultra-long bursts: a long quiescent time between the episodes ($\sim$1500 s), spectral evolution from hard initial pulse to significantly softer subsequent pulses, and a large fluence.
The duration of the second emission episode of GRB~080704 (following the quiescent time) is substantially longer that the duration of the first episode, which is a common feature of long GRBs. The existence of such long quiescent times may favor the dormant inner engine scenario over wind modulation models \cite{Drago2007}.

Without knowledge of the burst redshift it is not possible to determine the rest frame properties of the burst, but its large fluence and high $E_{peak}$ suggest a moderate z of $\sim$1--2 (for z=1 $E_{iso} \sim 1.2 \times10^{54}$ erg). In such a case, the rest frame burst duration of $\sim$1000 s and the rest frame quiescence time of $\sim$500 s can pose a problem for some central engine models in the framework of the collapsar scenario. Specifically such long durations might be hard to explain in the magnetar model of long GRBs.

The Konus-Wind experiment is supported by a Russian Space Agency contract and RFBR
grant 12-02-00032a. KH is grateful for IPN support under the following NASA grants:
NNX07AR71G (MESSENGER), NNX08AN23G (Swift), and NNX08AC90G (INTEGRAL).


\end{document}